\documentclass[12pt]{iopart}

\usepackage{iopams}  
\usepackage{graphicx}
\usepackage{color}

\begin{document}

\title[Shock mode]{Cyclic reformation of subcritical perpendicular fast magnetosonic shocks due to oblique Whistler waves}

\author{M. E. Dieckmann\footnote{Author to whom any correspondence should be addressed.}}
\address{Dept. of Science and Technology (ITN), Link\"oping University, Campus Norrk\"oping, SE-60174 Norrk\"oping, Sweden.}
\ead{mark.e.dieckmann@liu.se}

\author{L. Palodhi}
\address{Indian Institute of Technology Ropar, Department of Mathematics, Ropar 140001, Punjab, India}

\author{M.~François}
\address{CELIA, University of Bordeaux-CNRS-CEA, Talence 33405, France}

\author{D.~Folini}
\address{Centre de Recherche Astrophysique de Lyon (CRAL), ENS Lyon, Université de Lyon, Université de Lyon 1, CNRS, UMR5574, 46, allee d'Italie
69364 Lyon Cedex 07}

\author{R.~Walder}
\address{Centre de Recherche Astrophysique de Lyon (CRAL), ENS Lyon, Université de Lyon, Université de Lyon 1, CNRS, UMR5574, 46, allee d'Italie
69364 Lyon Cedex 07}

\vspace{10pt}
\begin{indented}
\item[]August 2025
\end{indented}

\begin{abstract}
The stability of subcritical perpendicular fast magnetosonic shocks, which are propagating at 1.7 times the fast magnetosonic speed, is investigated using two-dimensional PIC simulations. The plasma, composed of electrons and fully ionized nitrogen, is permeated by a uniform magnetic field oriented at 45$^\circ$ to the simulation plane normal. This configuration results in a diamagnetic current that sustains the shock’s magnetic ramp and is partially resolved within the simulation plane. The diamagnetic current drives an oblique lower-hybrid gradient drift instability within the ramp. This instability has been observed in magnetic reconnection experiments and studied in the framework of a Harris-type sheath in previous studies. It arises from a reactive coupling between the oblique Whistler wave, which is propagating backward in the electron rest frame, and the forward-propagating ion acoustic wave. Our simulations show that the magnetic component of this wave modulates the shock’s magnetic field, while the electrostatic ion density modulation forces the shock to collapse into a magnetic piston and then reform. The reformation is not forced by an external perturbation as in previous simulations but by the oblique Whistler wave.
\end{abstract}

%
\vspace{2pc}

\submitto{\NJP}
%
\maketitle
%
%

\section{Introduction}

Shock waves dissipate the directed flow energy of an inflowing upstream fluid. They convert it into heat and slow the fluid to a subsonic speed before it escapes downstream. Magnetohydrodynamic (MHD) shocks dissipate directed flow energy through binary collisions between the fluid particles. Binary collisions can maintain equal temperatures between the plasma species and keep their velocity distributions close to a Maxwellian, which is the thermal equilibrium distribution in nonrelativistic plasma. They also limit the wave spectrum to low frequencies. The ion acoustic wave is the only compressional fluid mode in unmagnetized plasma. A plasma with a uniform background magnetic field supports the ion acoustic wave and shear Alfv\'en wave with wave vectors parallel to the background magnetic field. The wave vectors of slow and fast magnetosonic waves point obliquely or perpendicular to the background magnetic field.

Shocks also exist in space and astrophysical plasmas~\cite{Treumann2009,Marcowith2016}, in which binary collisions between plasma particles rarely occur. Infrequent collisions cannot keep the plasma near its thermal equilibrium at the shock, leading to non-Maxwellian particle velocity distributions. Electrons decouple from the more massive ions and both species react only to macroscopic electric and magnetic fields. The different gyrofrequencies of electrons and ions and the different reaction to oscillating electric fields lead to a plasma response, which varies with the frequency and wavenumber of the driving fields. Some wave modes in collisionless plasma converge to one of the MHD wave branches if their frequency is below the lowest resonance frequency. In this case, the electrons and ions behave similarly and the one-fluid MHD theory becomes applicable. Electrons and ions can also have different mean velocities, which gives rise to electric currents. Collisions dissipate electric current via ohmic heating. In their absence, electromagnetic waves grow, disrupt the electric current, and heat the plasma.

Plasma, in which binary collisions are negligible, is referred to as collisionless. The shocks forming in such plasma are also called collisionless~\cite{Biskamp1973}. The dynamics of collisionless shocks and the nearby plasma can be quite different from that in collisional plasma. Collisionless shocks can, for example, accelerate a subset of particles to cosmic-ray energies, while the bulk of the plasma remains relatively cool~\cite{Drury1983}. Collisionless instabilities, which cannot develop in collisional plasma and depend for example on the angle the current forms with the magnetic field, can be strong enough to act back on the shock structure~\cite{Umeda2012,Muschietti2017,Denig2025}. The shock's structure and evolution therefore depend to a larger degree on the plasma composition and electromagnetic field distribution close to the shock than in collisional plasma. 

Unlike shocks in collisional fluids, which are characterized by jumps in field and plasma properties, collisionless shocks have a transition layer. The plasma density and magnetic field amplitude increase beyond their values upstream as they approach the transition layer of a perpendicular shock. Both go through an overshoot in the transition layer and decrease to the downstream values after they leave it. The shock surface is commonly defined as the layer in which the magnetic field amplitude overshoots.

The arguably best studied and most important collisionless shock is the Earth's bow shock. The Earth's magnetic field poses an obstacle to the solar wind, which is a fast-flowing magnetized plasma emanating from the solar corona. The bow shock slows the solar wind to a subsonic speed before the shocked solar wind enters its downstream region, known as the magnetosheath. Due to the curvature of the bow shock and the variability of the solar wind and the magnetic field it carries, we find a wide range of shock configurations. The perpendicular bow shock~\cite{BowShock} is characterized by a thin transition layer in which the plasma is heated and compressed. 

The magnetospheric multiscale mission (MMS) detected surface oscillations of the Earth's quasi-perpendicular bow shock~\cite{Johlander2016}. The shock speed in the upstream frame was about 4 times the fast magnetosonic speed and the shock was supercritical; perpendicular shocks become subcritical when their cross-shock electric potential is sufficiently strong to slow down the plasma, which requires a Mach number below 2.7~\cite{Marshall1955}. Supercritical shocks must dissipate some of the directed flow energy of the upstream plasma by other means. Typically this occurs through a dense beam of shock-reflected particles or through nonstationary shock dynamics in the shock rest frame. The oscillations observed by the MMS satellites at that occasion were identified as Alfv\'en waves that propagate along the magnetic field at the speed expected for the magnetic amplitude at the shock's surface. 

Collisionless plasma is modelled with particle-in-cell (PIC) codes, which use a kinetic approximation for electrons and ions, and
hybrid codes, which consider kinetic ions and an inertialess electron fluid. Lowe and Burgess~\cite{Lowe2003} studied supercritical perpendicular shocks with two-dimensional hybrid simulations and a magnetic field direction that was resolved by the simulation plane. Some inflowing upstream ions were reflected by the shock. Their free energy was released ahead of the shock causing the upstream plasma to become spatially nonuniform. As the shock ran into a nonuniform medium, it became corrugated. The magnetic field in the overshoot layer was deformed and magnetic tension led to Alfv\'en waves. A corrugation of the upstream plasma can thus lead to shock surface oscillations. 

Supercritical shocks are not stationary in time, and their dense beam of shock-reflected ions drives strong instabilities upstream of the shock~\cite{Muschietti2017,Lembege2009,Dimmock2019}. An example is the modified two-stream instability (MTSI)~\cite{TreumannBaumjohann} examined in Ref.~\cite{Umeda2012} and observed near the Earth's supercritical bow shock~\cite{Hull2020}. Its cause is the drift of the upstream ions and the shock-reflected ions relative to the electrons. The MTSI grows oblique Whistler waves, which propagate almost perpendicularly to the ambient magnetic field and turn into lower-hybrid waves for perpendicular propagation~\cite{Eliasson2025}. Another PIC simulation~\cite{Kobzar2021} showed that electron heating near a quasi-perpendicular shock drives oblique Whistler waves via the electron firehose instability. A comprehensive list of relevant instabilities that lead to the growth of such waves is given in Ref.~\cite{Instabilities}.

Subcritical perpendicular shocks~\cite{Forslund1970,Edmiston1984,Schaeffer2012,Niemann2014,Dieckmann2017} reflect fewer ions than their supercritical counterparts, which suppresses shock reformation due to the shock-reflected ion beam, leads to weaker instabilities upstream of the shock and, thus, less corrugation of the upstream plasma. Their transition layer is narrower than that of supercritical shocks and its oscillations can be studied with greater accuracy. At times, the solar wind speed is low enough to turn the bow shock into a subcritical one~\cite{Lugaz2016} and a better understanding of subcritical shocks may be relevant also for the Earth's bow shock.

Two-dimensional PIC simulations tracked a subcritical perpendicular shock. The upstream magnetic field direction was resolved by the simulation plane in~\cite{Dieckmann2023}. It was perpendicular to it in~\cite{Dieckmann2024}. Magnetic tension let the shock boundary oscillate and led to a periodic collapse and reformation of the shock~\cite{Dieckmann2025}. This reformation was much faster than that of supercritical shocks. The collapse of the shock in one boundary segment led to its reformation in the neighboring boundary segment. A corrugated upstream plasma in a PIC simulation therefore led to shock boundary oscillations. 

Instabilities near the overshoot can also cause shock surface oscillations. An example is the lower-hybrid drift instability, which is observed in the Earth's bow shock~\cite{Stasiewicz2020}, other shocks~\cite{Hanson2019} in the solar wind, and in the shock simulation in~\cite{Dieckmann2024}. 
The lower-hybrid drift instability drives waves that propagate strictly perpendicular to the magnetic field and the alignment of the magnetic field with the normal of a two-dimensional simulation plane favors their growth. It is important to understand how the spectrum of unstable waves is affected by a different magnetic field direction relative to the simulation plane normal. 

Here, a collisionless, subcritical perpendicular shock is studied with the particle-in-cell (PIC) code \emph{EPOCH}~\cite{Esirkepov2001,Arber2015}. The shock propagates into an upstream plasma that is permeated by a spatially uniform magnetic field that points orthogonally to the shock normal and at the angle 45$^\circ$ relative to the simulation plane normal. All initial conditions apart from the angle of 45$^\circ$ are identical to those in Refs.~\cite{Dieckmann2024,Dieckmann2025}. The lower-hybrid drift instability, which leads to electrostatic waves with no magnetic field component, does not grow; lower-hybrid waves are strongly damped unless they propagate almost perpendicularly to the magnetic field. Instead, oblique Whistler waves grow in the magnetic overshoot on the side facing the upstream plasma. Once the waves saturate they enforce shock reformation. This reformation does not require an external perturbation like in~\cite{Dieckmann2023,Dieckmann2025}.

It has been proposed that the electron drift associated with the diamagnetic current can lead to the growth of oblique Whistler waves~\cite{Biskamp, Lembege1992} in the transition layer of subcritical fast magnetosonic shocks but, to the best of our knowledge, the instability mechanism remained elusive. The simulations we perform bring forward evidence that the oblique lower-hybrid drift instability discussed in~\cite{Ji2005} is responsible for their growth. The instability becomes ineffective in our simulations if the wavenumber of the oblique Whistler wave approaches the largest unstable wavenumber predicted by the analytic work in~\cite{Ji2005}. For smaller wavenumbers, the oblique Whistler waves grow faster  in our simulations than predicted. The first aspect supports our claim that the oblique Whistler instability is responsible for the wave growth while the different growth rates can potentially be explained by electron heating at the shock.

Our paper is structured as follows. Section~2 summarizes the PIC algorithm and our initial conditions. We also discuss the plasma waves relevant to our work and the structure of the subcritical shock near its overshoot using data from PIC simulations in one spatial dimension. Section~3 examines with a PIC simulation in two spatial dimensions a shock, which moves into an unperturbed upstream plasma, and how it is modified by the oblique Whistler wave instability. Section~4 demonstrates that shock surface oscillations caused by a nonuniform upstream plasma are different from oscillations due to oblique Whistler waves and that they can suppress their growth. Section~5 discusses our results in the context of collisionless shocks in space- and laboratory plasmas. 

\section{Simulation setup and plasma conditions}

\subsection{Computational approach}

A PIC code describes each plasma species $i$ by a phase space density distribution $f_i(\mathbf{x},\mathbf{v},t)$. The position $\mathbf{x}$ and velocity $\mathbf{v}$ are independent coordinates. The distribution of each species $i$ is approximated by computational particles (CPs). Each CP with index $m$ has a position $\mathbf{x}_m$, velocity $\mathbf{v}_m$, and current density $\mathbf{j}_m \propto q_m\mathbf{v}_m$. Its charge $q_m$ and mass $m_m$ have the ratio $q_m/m_m$ that equals the $q_i/m_i$ of the represented species.

The macroscopic current density $\mathbf{J}$ is obtained by adding together the current density contributions $\mathbf{j}_m$ of all CPs. The current density $\mathbf{J}$ is coupled to the electric field $\mathbf{E}$ and the magnetic field $\mathbf{B}$ through Maxwell's equations
\begin{eqnarray}
    \nabla \times \mathbf{B} = \mu_0 \mathbf{J} + \mu_0 \epsilon_0 \frac{\partial \mathbf{E}}{\partial t},  \qquad 
     \nabla \times \mathbf{E} = - \frac{\partial \mathbf{B}}{\partial t}, \label{eq1} \\
     \nabla \cdot \textbf{B} = 0, \qquad \nabla \cdot  \textbf{E} = \rho/\epsilon_0,
     \label{eq2}
\end{eqnarray}
where $\mu_0$ and $\epsilon_0$ are the vacuum permeability and permittivity. Esirkepov's algorithm \cite{Esirkepov2001}, which is used by the \textit{EPOCH} code, fulfills the magnetic divergence law and Gauss' law with the charge density $\rho$ in Eqn.~\ref{eq2} as constraints. 

All components of $\mathbf{E}$, $\mathbf{B}$, and $\mathbf{J}$ are defined on a numerical grid with the cell size $\Delta_x$ and time step $\Delta_t$. From the position of the CPs in the numerical grid, the current contribution is obtained and then summed for all the CPs to give $\mathbf{J}$ which is then used to update the electromagnetic fields using Eqn.~\ref{eq1}. The new electromagnetic fields are interpolated to the positions of each CP and update its velocity and momentum using the relativistic Lorentz force equation
\begin{eqnarray}
\frac{d\mathbf{p}_m}{dt} = q_m(\mathbf{E} + \mathbf{v}_m \times \mathbf{B} ), \qquad \frac{d\mathbf{x}_m}{dt} = \mathbf{v}_m,
\label{eq3}
\end{eqnarray}
where $\mathbf{p}_m =  \gamma_m m_m\mathbf{v}_m$ is the relativistic momentum. A more detailed discussion of the \emph{EPOCH} code is given by Ref.~\cite{Arber2015}.

\subsection{Simulation setup and initial conditions}

A dense plasma expands into ambient plasma with conditions that are representative of laser-plasma experiments~\cite{Ahmed2013}. The ambient plasma's electrons have the number density $n_{e0} = 10^{21} \mathrm{m}^{-3}$, plasma frequency $\omega_{pe} = (e^2n_{e0}/\epsilon_0m_e)^{1/2} \approx 1.8 \times 10^{12} \; \mathrm{s}^{-1}$ ($e, m_e$: elementary charge and electron mass), and electron skin depth $\lambda_e = c/\omega_{pe}\approx 0.17$~mm ($c$: light speed). Space is expressed in units of $\lambda_e$.

\begin{figure*}
\includegraphics[width=\columnwidth]{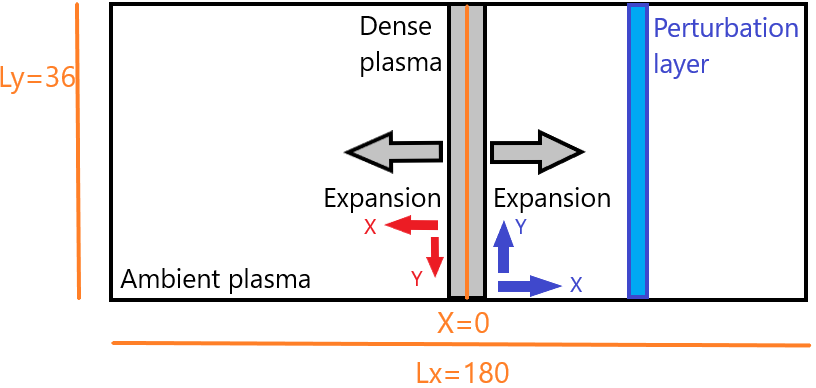}
\caption{Setup of the main simulation with the box size size $L_x \times L_y$ and periodic boundary conditions. Positions are expressed in units of the electron skin depth $\lambda_e$. The box is split in two. Each half has its own right-handed coordinate system with the z axis pointing up. A slab of dense plasma with thickness $6\lambda_e$ is centered on $x=0$ and surrounded by ambient plasma. The plasma conditions are uniform within the dense slab and within the ambient plasma outside the perturbation layer. A spatially uniform magnetic field $\mathbf{B}_0$ forms a right angle with x and is rotated relative to z by the angle $\theta=45^\circ$. Its $B_y$ component has opposite signs in the red and blue coordinate systems because their y-axes point in opposite directions. The dense plasma expands into the directions marked by the gray arrows. It expands into uniform plasma to the left and encounters a perturbation layer to the right.}
\label{figure01}
\end{figure*}

Figure \ref{figure01} sketches the setup of the 2D simulation box with the side lengths $L_x=180\lambda_e$ and $L_y=36\lambda_e$, which are resolved by 9000 and 1800 grid cells, respectively. The box is split in two equal halves at $x=0$. Each half has its own right-handed coordinate system with $0 \le x \le 90\lambda_e$ and $0 \le y \le 36\lambda_e$. The dense plasma expands in the direction of increasing x values. The background magnetic field $\mathbf{B}_0$ is spatially uniform across the box. Its orientation is $\mathbf{B}_0=(0,B_0/\sqrt{2},B_0/\sqrt{2})$ in the blue coordinate system and $\mathbf{B}_0=(0,-B_0/\sqrt{2},B_0/\sqrt{2})$ in the red coordinate system. Its amplitude $B_0 = 0.85$T gives $\omega_{ce}/\omega_{pe}=0.084$, where $\omega_{ce} = eB_0/m_e$ is the electron gyrofrequency. The boundary between the ambient and dense plasma is located at $x=3\lambda_e$. The perturbation layer covers $8.9\lambda_e \le x \le 20.8\lambda_e$ in the right half of the box in Fig.~\ref{figure01}. 

The electrons in the ambient plasma have the temperature $T_{e0} = 1$ keV and thermal pressure $P_{th} = n_{e0}k_B T_{e0}$ ($k_B$: Boltzmann constant) giving $\beta=P_{th}/P_B = 0.56$ for a magnetic pressure $P_B=B_0^2/(2\mu_0)$. The thermal speed, thermal gyroradius $r_g =v_{th,e}/\omega_{ce}$, and Debye length of the electrons are $v_{th,e} = (k_B T_{e0}/m_e)^{1/2} \approx 1.3 \times 10^{7}$~m/s, $r_{ge} = 0.5 \lambda_e$, and $\lambda_D = 0.044 \lambda_e$. 

Nitrogen with the ionization state $Z=7$ and mass $m_i \approx 2.6 \times 10^{4} m_e$ is the carrier of positive charge. Its number density outside of the perturbation layer and dense plasma is $n_{i0} = n_{e0}/7$. Its plasma frequency $\omega_{pi} = (Z^2e^2n_{i0} / \epsilon_0m_i)^{1/2}$ and gyro-frequency $\omega_{ci}=ZeB_0/m_i$ are $\omega_{pi}\approx 3 \times 10^{10}$ s$^{-1}$ and $\omega_{ci}=1.4 \times 10^{-3}\omega_{pi}$. The ion temperature is $T_{i0} = T_{e0}/5 = 0.2$~keV and thermal speed $v_{th,i}={(k_BT_i/m_i)}^{1/2}$ give the thermal gyroradius $r_{gi}\approx 5.3\lambda_e$. Their lower temperature and number density give the ions the thermal pressure $P_{th}/35$. The ion skin depth $\lambda_i \approx 60\lambda_e$, which equals 1~cm for our plasma parameters.  

The dense plasma has the same ion composition and temperature as the ambient plasma. Its density is 60 times higher and the electron temperature is $1.5T_{e0}$. The thermal pressure, which is about $90 P_{th}$, lets the dense plasma expand along increasing x in both coordinate systems depicted in Fig.~\ref{figure01} driving shocks. The shock expanding to the left propagates into a uniform plasma. The shock that moves to the right passes through a perturbation layer, where the number density of the mobile ions varies as $n_{i0,mob}/n_{i0} = 0.7 + 0.3\sin{(2\pi y/L_y)}$, while that of the electrons is kept at $n_{e0}$. Their charge difference is canceled by an immobile charge distribution and the initial electric field is set to zero. Electrons and ions are represented by 25 CPs per cell each.  The simulation resolves $t_{sim}=10^{-8}$s  by $1.365 \times 10^6$ time steps $\Delta_t$. The ions are practically unmagnetized because $\omega_{ci}t_{sim}\approx 0.4$. This setup is that of the main simulation. 

Simulations 1 and 2 are 1D simulations used for identifying the wavemodes in the ambient plasma with the aforementioned plasma parameters. Their simulation direction is aligned with x. In both cases, the length of the simulation box with periodic boundaries is $2L_x$, which is resolved by 9000 grid cells. The electrons and the fully ionized nitrogen are represented by 100 CPs per cell, respectively. Both simulations differ only in the direction of $\mathbf{B}_0$. Simulation 1 aligns it with the z axis. Simulation 2 rotates it in the x-z plane by $\theta = 45^\circ$ relative to the z axis. 

Simulations~3-5 model a dense plasma expanding into a uniform ambient plasma like in the red coordinate system in Fig.~\ref{figure01}. Simulation~3 is a 1D simulation that resolves a slice along the x axis. It uses 2000 CPs per cell for electrons and ions, respectively. Simulations~4 and 5 are 2D simulations that use the same initial conditions and number of CPs per cell as the main simulation. Their box lengths along y are $3L_y/4$ and $L_y/2$, respectively. They track the plasma expansion until $0.6\, t_{sim}$. Table~\ref{table1} summarizes the key features of these simulations and their purpose.

\begin{table}
\begin{tabular}{|c|c|c|c|}
\hline
Label & Box dimension & Purpose & B-field direction \\
\hline
Simulation~1 & 1D: Length: $2L_x$ & Dispersion relation &  $\mathbf{B}_0 = (0, 0, B_0)$ \\ 
Simulation~2 & 1D: Length: $2L_x$ & Dispersion relation &  $\mathbf{B}_0 = (B_0, 0, B_0)/\sqrt{2}$ \\ 
Simulation~3 & 1D: Length: $L_x$ & Shock simulation & $\mathbf{B}_0 = (0,0,B_0)$ \\
Simulation~4 & 2D: Area: $L_x \times 3L_y /4$ & Shock simulation & $\mathbf{B}_0 = (0, B_0, B_0)/\sqrt{2}$ \\
Simulation~5 & 2D: Area: $L_x \times L_y /2$ & Shock simulation & $\mathbf{B}_0 = (0, B_0, B_0)/\sqrt{2}$ \\
Main simulation & 2D: Area: $L_x \times L_y $ & Shock simulation & $\mathbf{B}_0 = (0, B_0, B_0)/\sqrt{2}$ \\
\hline
\end{tabular}
\caption{Purpose and geometry of the 6 simulations presented here. Simulations with the purpose "Dispersion relation" model a spatially uniform plasma with the parameters of the ambient plasma in Section 2.2. These simulations compute the electromagnetic fluctuation spectrum, which allows us to identify the relevant wave modes in the plasma. Simulations with the purpose "Shock simulation" model a plasma with a density jump 60 between the dense and ambient plasma. The plasma parameters of the dense plasma are those discussed in Section~2.2. Shock simulations track the shocks and investigate the plasma conditions near them. The magnetic field orientation for the main simulation is that in the blue coordinate system in Fig.~\ref{figure01}.}
\label{table1}
\end{table}

Our initial conditions differ from those in the previous simulations discussed in Refs.~\cite{Dieckmann2023,Dieckmann2024,Dieckmann2025} only in the direction of the background magnetic field $\mathbf{B}_0$ relative to the normal direction of the simulation plane. Like in the previous simulations, $\mathbf{B}_0$ is orthogonal to the average shock normal that points along $x$. In the simulation in Ref.~\cite{Dieckmann2023}, $\mathbf{B}_0$ pointed along $y$. It pointed along $z$ in the simulation in Ref.~\cite{Dieckmann2024}. Both simulations were compared in~\cite{Dieckmann2025}. The orientation of $\mathbf{B}_0$ determines the direction, in which the electrons drift in the shock transition layer. A spatially varying magnetic field is tied to an electric current. This electric current is provided by electrons that drift perpendicularly to the shock normal and to the magnetic field direction. In the simulation~\cite{Dieckmann2023}, the electrons drift along the numerically unresolved z-direction. Unstable waves near the shock have a wave vector which is parallel to the electron drift direction and were not resolved by that 2D simulation. Current dissipation was absent in that simulation. The shock collapsed and reformed due to magnetic tension and resulted in a shock boundary oscillation that did not propagate along the boundary. 

Aligning the magnetic field with the numerically unresolved z-direction in~\cite{Dieckmann2024} removed magnetic tension, which suppressed shock boundary oscillations due to periodic shock collapse and reformation. The electron drift direction along $y$ was numerically resolved in this simulation and gave rise to electron-cyclotron drift instabilities ahead of the shock transition layer and lower-hybrid drift instabilities behind it. Electron-cyclotron waves and ion-cyclotron waves, which propagate perpendicular to the magnetic field, are electrostatic and have no magnetic field component. They can only modulate the magnetic field by compressing a magnetized plasma. 
 
In the simulation setup discussed above, $\mathbf{B}_0$ has equally strong components along $y$ and $z$. A magnetic field component along $y$, which is weaker than that in~\cite{Dieckmann2023}, results in a weaker magnetic tension force if the magnetic field is deformed. The magnetic field component along $z$ leads to a numerically resolved $y$ component of the electron drift direction, which can give rise to wave instabilities. Since electrostatic electron-cyclotron waves and ion-cyclotron waves are heavily damped if their wavevector and $\mathbf{B}_0$ form an angle of $45^\circ$, we expect that different instabilities grow.

\subsection{Relevant plasma waves}

The ambient plasma supports several wave modes. The fast magnetosonic (FMS) wave has low frequencies $\omega$ and wave numbers $k = 2\pi / \lambda$. Magnetohydrodynamic (MHD) theory applies for $\lambda \gg r_{ge}, r_{gi}$. The dispersion relation
\begin{equation}
\omega^{2} = \frac{k^2}{2} 
\left (
(c_s^2+v_A^2) + \left [
(v_A^2-c_s^2)^2+4v_A^2c_s^2\frac{k_\perp^2}{k^2}
\right ]^{1/2}
\right )
\label{fullvfms}
\end{equation}
describes FMS waves propagating at an angle relative to $\mathbf{B}_0$ set by the wave vector $\mathbf{k}$ and its components $k_\parallel$ and $k_\perp$ parallel and perpendicular to $\mathbf{B}_0$. The Alfv\'en speed and ion acoustic speed are $v_A = B_0 /(\mu_0 m_i n_{i0})^{1/2}$ and $c_s = {(k_B(Z\gamma_e T_{e0}+\gamma_iT_{i0})/m_i)}^{1/2}$. The adiabatic constant $\gamma_i$ for the fully ionized nitrogen with $Z=7$ is $\gamma_i=3$. If $B_0=0$ and $T_{e0}\gg T_{i0}$, the electrons are isothermal with $\gamma_e=1$~\cite{Noise}. We set it to $\gamma_e=5/3$ to make it consistent with our previous work giving $c_s=2.9 \times 10^5$~m/s and $v_A=4.15 \times 10^5$~m/s in the ambient plasma outside the perturbation layer. 

For wave propagation perpendicular to $\mathbf{B}_0$ with $k_\parallel=0$, Eqn.~\ref{fullvfms} becomes $\omega / k = v_{fms}$ with the FMS speed $v_{fms}=(c_s^2+v_A^2)^{1/2}\approx 5\times 10^5$ m/s. The MHD approximation breaks down once the wave frequency gets close to its resonance at the lower-hybrid frequency $\omega_{lh}=((\omega_{ci}\omega_{ce})^{-1} + \omega_{pi}^{-2})^{-1/2}$. The FMS mode, which is electromagnetic for low $k$, acquires an electrostatic component $\mathbf{E}\parallel \mathbf{k}$. The electrostatic approximation of the linear wave dispersion relation for lower-hybrid waves is 
\begin{equation}
\omega = \left ( 3v_{th,i}^2k^2 + \frac{\omega_{pi}^2(\omega_{ce}^2+v_{th,e}^2k^2)}{\omega_{pe}^2+\omega_{ce}^2 + v_{th,e}^2k^2}  \right )^{1/2}.
\label{Lower-Hybrid}
\end{equation}

\begin{figure*}
\includegraphics[width=\columnwidth]{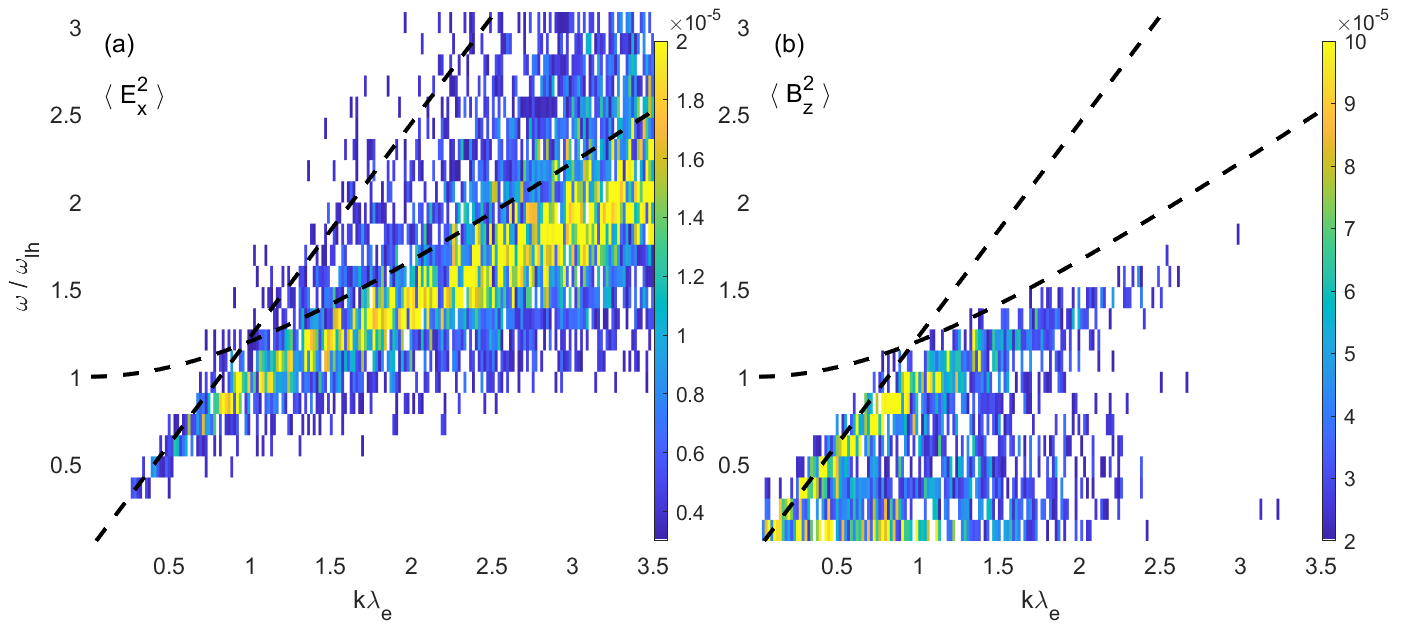}
\caption{The power spectra of $\langle E_x^2 \rangle$ and $\langle B_z^2 \rangle$ for wave propagation perpendicular to the background magnetic $\mathbf{B}_0$ (simulation~1) are shown in~(a) and~(b), respectively. The power in both panels is normalized to the values of $c^2\langle B_z^2\rangle$ and $\langle B_z^2 \rangle$ at $k,\omega=0$, respectively. The dashed straight line, which is the electromagnetic approximation of the FMS/lower-hybrid wave branch, is $\omega_{fms}=v_{fms}k$. The dashed curve is the solution of Eqn.~\ref{Lower-Hybrid}, which is the electrostatic approximation of the FMS/lower-hybrid wave branch. The fluctuation spectrum demonstrates a gradual change from the electromagnetic approximation to the electrostatic one near $\omega_{lh}$}.
\label{figure02}
\end{figure*}

Neither the electrostatic approximation for lower-hybrid waves nor the electromagnetic approximation for FMS waves is exact near the lower-hybrid frequency $\omega_{lh}$. We can test the accuracy of these approximations by comparing their solutions of the linear wave dispersion relation to the noise spectrum computed by a PIC simulation. The power spectrum of field fluctuations peaks at values of $k, \omega$ close to those of plasma Eigenmodes. Simulation~1 modelled the uniform ambient plasma, resolved the x direction, and aligned $\mathbf{B}_0$ with the z axis. Multiplying the Fourier transforms of $E_x,B_y$ in space and time with their complex conjugates gives us the power spectra $\langle E_x^2 \rangle$ and $\langle B_z^2 \rangle$ of the fluctuations. Figure~\ref{figure02} compares them with the solutions of the linear dispersion relations. In the limit $k\rightarrow 0$, the noise becomes magnetic and converges to the FMS mode~\cite{Eliasson2025}. Dispersive effects leading to a decreasing phase speed of the FMS mode with increasing $k$ set in at around 0.6$\omega_{lh}$.
With increasing $k\lambda_e>1$, the noise becomes electrostatic and follows Eqn.~\ref{Lower-Hybrid} albeit at a somewhat lower frequency. 
The lower frequency of the noise peak in Fig.~\ref{figure02}(a) might be caused by effects, which are not taken into account by the warm fluid approximation leading to Eqn.~\ref{Lower-Hybrid}. The most likely being electrostatic ion-cyclotron waves, which are linearly undamped for propagation perpendicular to the magnetic field. Ion-cyclotron modes exist in collisionless plasma but are not captured by MHD theory.

Our aim is to study shock boundary oscillations. Simulations 4, 5, and the main simulation (See Table~\ref{table1}) track the expansion of shocks along the x-direction. The average normal of these shocks is thus parallel to $x$. These 2D simulations also resolve the y-direction and, thus, all waves with wavevectors in the x-y plane. The background magnetic field $\mathbf{B}_0$ in these simulations points orthogonally to the average shock normal and the shock is thus mediated by FMS/lower-hybrid waves. The magnetic field $\mathbf{B}_0$ is tilted by $45^\circ$ degrees relative to $z$. The resolved waves, which propagate along the shock boundary in the 2D simulation plane, thus propagate at an angle $45^\circ$ relative to $\mathbf{B}_0$. These waves become oblique FMS waves for low $k$. Their wave vectors fullfill $k_\perp^2/k^2=1/2$, which reduces the phase speed of the oblique FMS waves to $0.93v_{fms}$. Oblique FMS waves are compressive only for propagation angles $\theta \gtrsim \theta_c$ with $cos^2\theta_c = c_s^2/v_{fms}^2$ or  $\theta_c \approx 55^\circ$ in the ambient plasma. Oblique FMS waves that propagate along the shock boundary can thus not modulate the ion density. 

At large $k$, $k_\perp \neq 0$, and $\omega > \omega_{ci}$ we find the oblique Whistler waves fullfilling~\cite{Artemyev2016}
\begin{equation}
\omega^2 = \frac{\omega_{ce}^2k_\perp^2/k^2}{(1+(\omega_{pe}/kc)^2)^2}+
\frac{\omega_{lh}^2}{1+(\omega_{pe}/(kc))^2}. 
\label{whistler}
\end{equation}
A comparison of the solution of the linear dispersion relation for oblique Whistler waves propagating at the angle $45^\circ$ relative to $\mathbf{B}_0$ with the power spectrum of the PIC simulation noise sheds light onto how accurate Eqn.~\ref{whistler} is and if other wave modes exist in this frequency and wavenumber range.

Simulation~2 provides the power spectra $\langle E_x^2 \rangle$ and $\langle B_z^2 \rangle$ for $\theta = 45^\circ$.
\begin{figure*}
\includegraphics[width=\columnwidth]{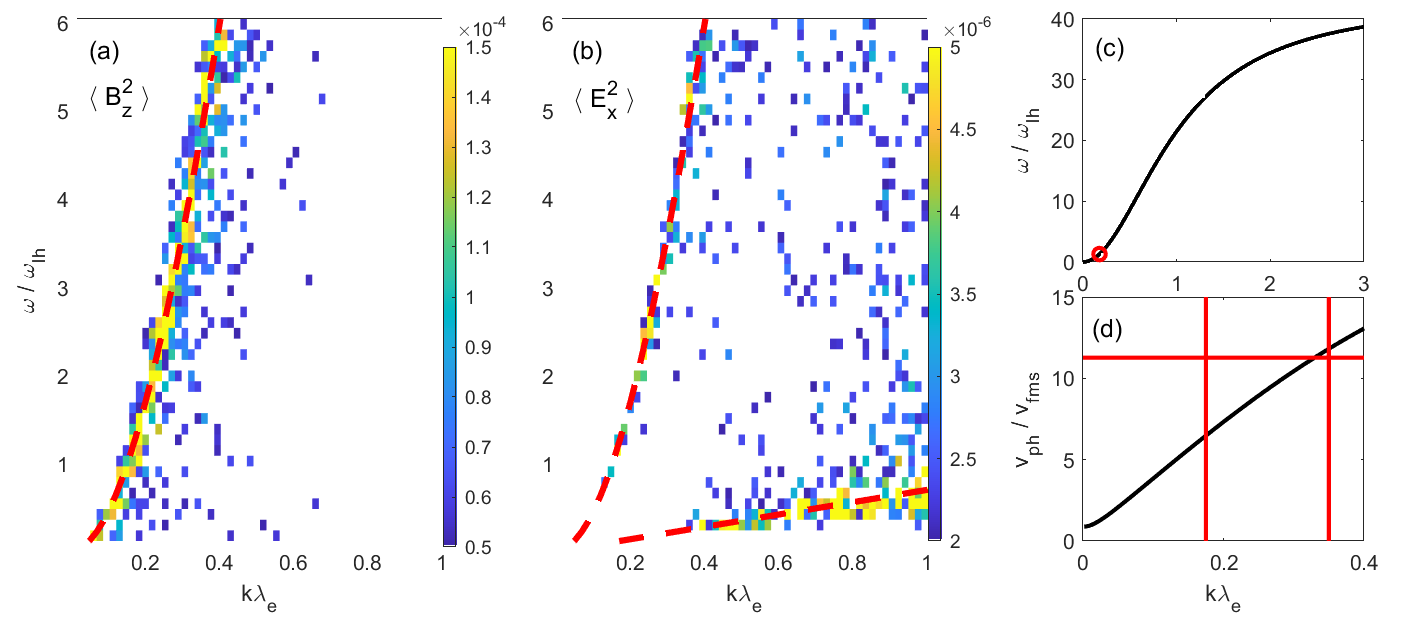}
\caption{Waves propagating at the angle $\theta= 45^\circ$ relative to $\mathbf{B}_0$ (Simulation~2). Panels~(a,~b) show $\langle B_z^2 \rangle$ and $\langle E_x^2 \rangle$, respectively. The power spectra are normalized to $\langle B_z^2 \rangle$ and $c^2\langle B_z^2 \rangle$ at $k,\omega=0$. The dashed red curves show the solution of Eqn.~\ref{whistler} and the red dashed line in~(b) shows $\omega = c_s k$ of the ion acoustic wave. Panel~(c) shows the dispersion relation extending to large $k$ and $\omega$. The red circle corresponds to $k=2\pi / L_y$ and $\omega = 1.28\omega_{lh}$. Panel~(d) compares the phase speed $v_{ph}=\omega / k$ of the oblique Whistler wave with the speed $5.3 \times 10^6$ m/s (horizontal line). The vertical lines show $k = 2\pi/L_y$ and $4\pi/L_y$.}
\label{figure03}
\end{figure*}
Figures~\ref{figure03}(a,~b) show the solution of Eqn.~\ref{whistler} for oblique Whistler waves with $\theta = 45^\circ$ relative to $\mathbf{B}_0$ together with the fluctuation spectra. The frequency of oblique Whistler waves increases faster than that of the FMS/lower-hybrid wave branch shown in Fig.~\ref{figure02}. The magnetic fluctuations reach their peak power close to the $\omega, k$ of oblique Whistler waves. Figure~\ref{figure03}(b) shows noise that follows their branch at high $\omega$ evidencing an electrostatic component. No electrostatic noise is visible at low frequencies, which is probably a consequence of the oblique FMS mode not being a compressive mode. Additional electrostatic noise is concentrated near the dispersion relation $\omega = c_s k$.

Figure~\ref{figure03}(c) shows the solution of the linear dispersion relation for oblique Whistler waves for a wide interval in $k$ and $\omega$. The red circle marks $k=2\pi / L_y$ and the frequency of the oblique Whistler wave for this wave number. Figure~\ref{figure03}(d) shows the phase speed of the oblique Whistler wave. It exceeds $v_{fms}$ by far for $k=2\pi/L_y$ and $4\pi / L_y$. The phase speed for $k=2\pi \lambda_e/L_y$ is $0.23v_{th,e}$.


\subsection{Shock formation and structure}

In collisional plasma, Coulomb collisions between plasma particles dissipate the directed flow energy of the upstream plasma that crosses the shock. The dissipation heats the plasma. The shock balances the thermal pressure of the heated plasma downstream of the shock with the ram pressure of the inflowing cooler upstream plasma. The shock thickness is comparable to the mean-free path of plasma particles, which is taken to be infinitesimally small in MHD theory. 

In the absence of binary collisions, the mean-free path is larger than the scales of interest. Other mechanisms must therefore be at work when collisionless shocks form. 
We distinguish between dissipative and dispersive mechanisms. Both have in common that a shock forms by the steepening of a large-amplitude wave. Wave steepening, which is discussed in Refs.~\cite{Dawson1958,Shukla2004}, couples energy from waves with a long wavelength to short waves. As long as all waves are undamped and have the same phase speed, the wave front continues to steepen. Eventually, the waves that form the wave front reach a wavelength where either wave damping or dispersive effects become important~\cite{Margaux}.

Wave damping in collisionless plasma is often caused by resonant interactions between the electromagnetic wave field and the electrons and ions, which accelerate and decelerate particles. A plasma in thermal equilibrium has a Maxwellian velocity distribution, which decreases with increasing velocity. Resonances thus accelerate more slow particles than they decelerate fast particles. On average, the wave loses energy to the particles. Resonant damping known as Landau damping usually gets stronger with increasing wavenumbers. Shock steepening, which couples energy from long to short waves, is in this case halted by Landau damping.

The shocks in our simulations are mediated by FMS/lower-hybrid waves. Lower-hybrid waves, which propagate perpendicular to $\mathbf{B}_0$, are linearly undamped because the magnetic field restricts the electron mobility along the wave vector thereby suppressing Landau damping. Unless the lower-hybrid waves are strong enough to trigger nonlinear processes, the shocks in our simulations cannot be dissipative.

Figure~\ref{figure02}(a) shows that the coupled FMS/lower-hybrid wave branch is dispersive. Its phase speed decreases with increasing $k$. Shorter waves fall behind the wave front~\cite{Dieckmann2017}. Shorter waves outrun the wave front if the phase speed increases with $k$, like in the case of the oblique FMS/Whistler waves at frequencies $\approx \omega_{lh}$~\cite{Moreno2018}. In either case, the short waves leave the vicinity of the shock front leaving behind the longer waves, which are less dispersive. The loss of short waves thus inhibits further steepening of the wave front. The thickness of the shock transition layer is in this case comparable to the wavelength, at which dispersive effects become important. Figure~\ref{figure02} shows that dispersion sets in at $k\lambda_e \approx 1$ giving a wavelength $\sim 2\pi \lambda_e$.

\begin{figure*}
\includegraphics[width=\columnwidth]{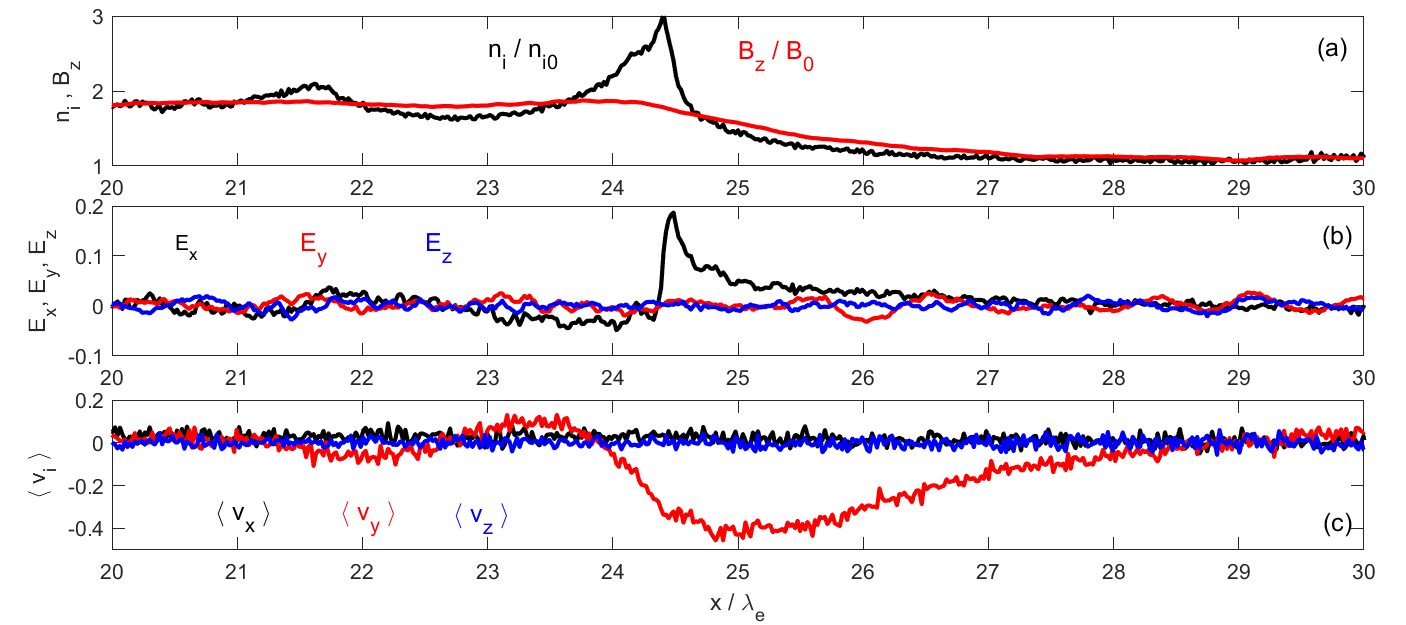}
\caption{The profile of a perpendicular subcritical shock at the time $\omega_{lh}t=12$. Panel~(a) shows the ion density in units of $n_{i0}$ (black) and $B_z/B_0$; the background magnetic field pointed along the z axis in simulation~3. Panel~(b) shows the electric field amplitudes $E_x$, $E_y$, and $E_z$ in units of $cB_0$. Panel~(c) shows the electron mean velocity $\langle v_i \rangle$ along the directions $i=x, y, z$ expressed in units of $v_{th,e}$.}
\label{figure04}
\end{figure*}

An observed width $\lesssim 2\pi \lambda_e$ of the shock transition layer would support our claim that the FMS/lower-hybrid shocks, which propagate perpendicularly to $\mathbf{B}_0$, are dispersive for our plasma parameters. One purpose of simulation~3 is to determine this width. Simulation~3 lets plasma expand along x into the ambient plasma with a magnetic field that is aligned with z. Figure~\ref{figure04} shows the ion and field distributions at $\omega_{lh}t=12$.  
Figure~\ref{figure04}(a) shows the density overshoot at $x/\lambda_e \approx 24.5$. The density change ahead of it induces a thermoelectric field along x in Fig.~\ref{figure04}(b). A weaker negative $E_x$ field is induced behind the density overshoot in $23 \le x / \lambda_e \le 24.2$. A second density maximum located at $x/\lambda_e \approx 21.5$ demonstrates that the wave mode that mediates the shock has the wavelength $\lambda \approx 3\lambda_e$ (peak-to-peak distance). This $\lambda$ corresponds to the wavenumber $k\lambda_e =2\pi / 3 > 1$ placing it into the wavenumber range where dispersive effects become important for the FMS/lower-hybrid wave branch discussed in Section~2.3. The magnetic overshoot reaches its peak value at $x/\lambda_e \approx 24$ and extends up to $x/\lambda_e \approx 27.5$. A second weaker maximum is located at $x/\lambda_e \approx 21.5$. 

Figure~\ref{figure04}(c) shows the mean velocities of the electrons along the three cartesian directions. Those along x and z hardly change. The electron drift velocity modulus along y reaches $|v_D|= 0.4v_{th,e}$ or $|v_D|=5.3\times 10^6$ m/s near $x/\lambda_e \approx 25$ and oscillates with a lower amplitude in the interval $21.5 \le x/\lambda_e \le 24$. According to Fig.~\ref{figure03}(d), $|v_D|$ exceeds the phase speed of oblique Whistler waves up to the wavenumber $\approx 4\pi/L_y$.  

Electrons drifting with the speed $v_D$ in a spatially varying magnetic field give rise to a gradient-drift instability. Its unstable wave spectrum was derived in~\cite{Ji2005} for a Harris sheath-like distribution, in which the changing magnetic pressure is balanced by a nonuniform thermal pressure. 

The magnetic pressure change in the overshoot of a subcritical FMS shock like that in simulation~3 is balanced by the ram pressure of the inflowing upstream plasma. Given that the ion flow speed $1.7v_{fms}$ across the subcritical shock in Fig.~\ref{figure04} is small compared to $|v_D|\approx 10.6v_{fms}$, some of the analytic results from~\cite{Ji2005} should also hold here. 

We summarize aspects of the instability~\cite{Ji2005}, which are relevant for our work. The authors employ a magnetized fluid model for the electrons and consider the ions to be unmagnetized and drifting relative to the electrons. The instability can thus be reduced to the interaction between oblique Whistler waves and Doppler-shifted ion acoustic waves. The magnetic field will change the ion acoustic wave into a magnetosonic wave, which is likely to have some effects on the instability. At the same time the authors of~\cite{Ji2005} note that their idealized model leads to a better and more intuitive understanding of the instability mechanism. It also simplifies the derivation of the range of unstable wave numbers and the exponential growth rates of the oblique Whister waves. The gyrokinetic approximation Ref.~\cite{Ji2005} uses for electrons is valid only for wavelengths that are significantly larger than the electron's thermal gyroradius or $k\lambda_e \ll 2\pi \lambda_e / r_g \approx 4\pi$. Instability arises from a reactive coupling between the ion acoustic wave and the Doppler-shifted backward-propagating oblique Whistler wave. 

A reactive instability is one in which the bulk of the plasma particles participate. They differ from resonant instabilities, where waves with the phase velocity $\omega/k$ interact only with plasma particles with a comparable speed. Our main simulation discussed in section~3 shows that the phase speed of the oblique Whistler wave in the ion's rest frame exceeds by far the ion's thermal speed thus excluding wave-ion resonances. Resonant instabilities also require particles to maintain their phase relation with the wave for a sufficiently long time. In our shock simulations, electrons and ions remain near the shock only for a short time before they escape downstream. It is thus unlikely that a resonant instability can develop near our shock. The instability observed in the main simulation should thus be reactive like the one discussed in the analytic work by~\cite{Ji2005}.


According to~\cite{Ji2005}, oblique Whistler waves grow in a wavenumber interval $k_{min} \le k \le k_{max}$, which depends on $v_D$, the wave propagation angle $\theta$, and the thermal-to-magnetic pressure ratios $\beta_e$ of the electrons and $\beta_i$ of the ions. Our ambient plasma has $\beta_e \approx 0.56$, $\beta_i = \beta_e / 35$, $\theta = 45^\circ$, and simulation~3 gives $|v_D|=10.6v_A$. The case $\beta_e = \beta_i = 1$ and $|v_D|=10v_A$ is covered in Ref.~\cite{Ji2005} giving $k_{min}\lambda_i \approx 5$ and $k_{max} \lambda_i\approx 15$, which corresponds to $k_{min}\lambda_e\approx 0.08$ and $k_{max}\lambda_e \approx 0.25$. 

Based on the values of $k_{min}$ and $k_{max}$, the oblique gradient-drift instability lets waves grow with wavelengths between $\lambda \approx 80\lambda_e$ and $\lambda \approx 25\lambda_e$. We expect the growth of an oblique Whistler wave with the wavelength $\lambda = L_y$, while its first harmonic with $\lambda \approx L_y/2$ that is resolved in the main simulation should be stable. The warm fluid model in~\cite{Ji2005} predicts a peak growth rate of the wave $\approx 10\omega_{ci}$ or $0.16\omega_{lh}$.   

In contrast to the Harris-type sheath in~\cite{Ji2005}, a perpendicular shock compresses the magnetic field. Magnetic field compression heats electrons perpendicularly to $\mathbf{B}_0$ but not parallel to it. Whistler wave instabilities tend to grow faster with increasing ratios between the electron temperature perpendicular to the magnetic field and parallel to it. It is thus likely that the oblique Whistler waves will grow faster in our shock simulation than in the Harris-type configuration considered by~\cite{Ji2005}. Equation~\ref{whistler} shows that the solution of the dispersion relation of the oblique Whistler wave and, thus, its phase speed do not depend on the electron temperature. It is thus likely that the range of unstable wave numbers derived for a Harris sheet-type configuration also applies to perpendicular shocks.

\section{Shock moving into unperturbed upstream plasma}

We present data from the main simulation and consider the shock, which propagates into the uniform ambient plasma to the left in Fig.~\ref{figure01}. In this section, we show $-B_y$ instead of $B_y$ so that we can use the same color scale for $B_y$ and $B_z$ and we refer with $B_y$ to $-B_y$. The magnetic $B_x$ exceeds noise levels only where magnetic field lines are deformed fullfilling the magnetic divergence law.

\begin{figure*}
\includegraphics[width=\columnwidth]{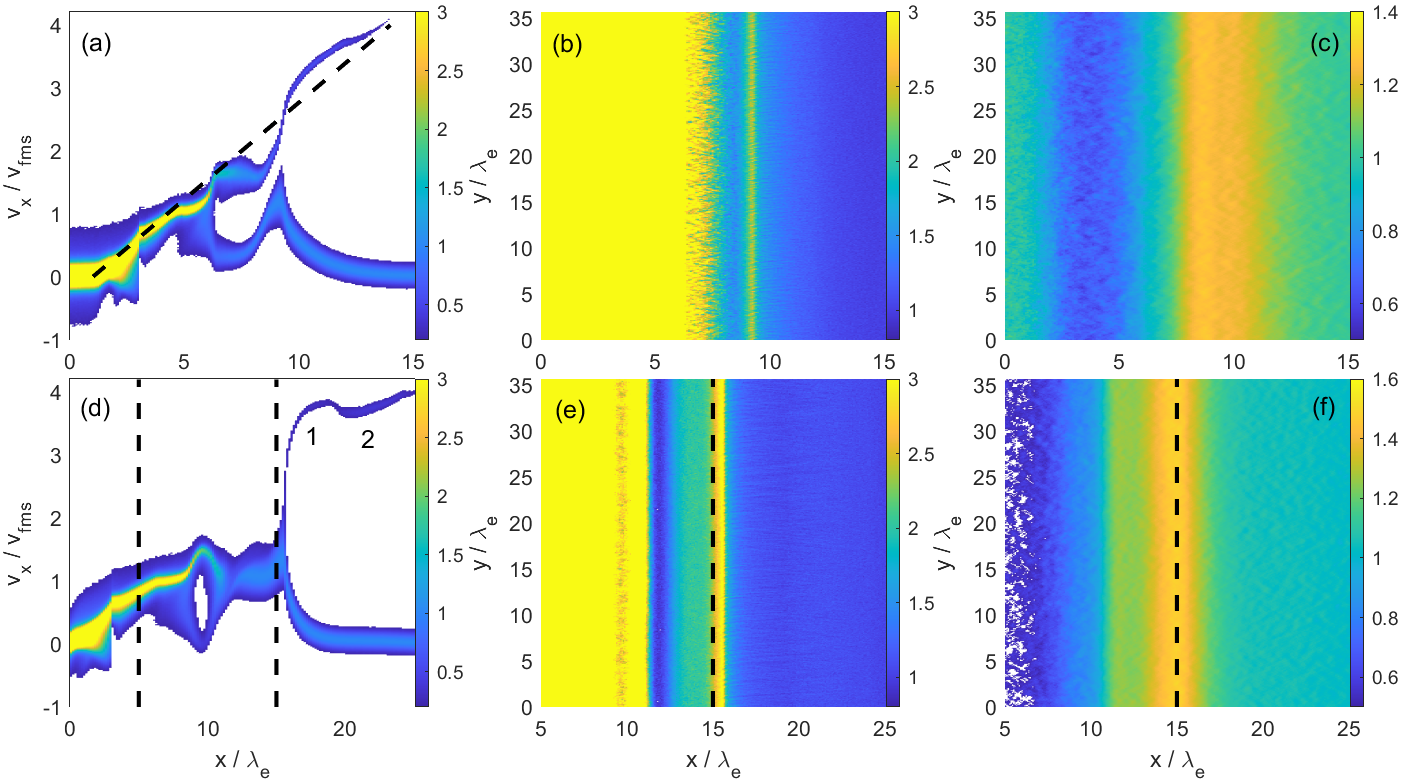}
\caption{Ion and magnetic field distributions at $\omega_{lh}t=2.9$ (upper row) and $\omega_{lh}t=5.8$ (lower row): Panels~(a,~d) show the square root of the phase space density, which was normalized to the peak value upstream. The dashed line in~(a) follows the profile of the rarefaction wave and those in~(d) show $x/\lambda_e = 5$ and 15. The ions above 1 correspond to shock-reflected ions and those above 2 to rarefaction wave ions. Panels~(b,~e) show the ion density normalized to the upstream value $n_{i0}$ and clamped to the value 3. Panels~(c,~f) show the modulus $| \mathbf{B}| /B_0$ of the magnetic field. The vertical dashed lines in~(e,~f) show $x/\lambda_e = 15$.}
\label{figure05}
\end{figure*}

Figure~\ref{figure05} shows the early shock evolution. The ion phase space density distribution in Fig.~\ref{figure05}(a) sampled at $\omega_{lh}t=2.9$ shows the rarefaction wave, which is formed by the accelerating ions of the dense plasma and marked by the diagonal dashed line. The ions are accelerated to a speed larger than $v_{fms}$ at $x = 5\lambda_e$ resulting in supersonic bulk flow. A FMS wave grows near $x \approx 9\lambda_e$. It has already steepened and is about to break. The ion density $n_i(x,y)$ in Fig.~\ref{figure05}(b) shows the expanding dense plasma at low values of x. The FMS wave growing in the ambient plasma gives rise to the narrow density maximum. The expanding rarefaction wave pushes out the magnetic field in Fig.~\ref{figure05}(c) creating the magnetic cavity around $x\approx 4\lambda_e$ and a pileup at larger values of x. The width along x indicates that the magnetic oscillation is connected to the rarefaction wave at $\omega_{lh}t=2.9$ and not to the growing FMS wave. 

At $\omega_{lh}t = 5.8$, the breaking FMS wave led to the growth of a phase space vortex near $x = 10\lambda_e$ and to the FMS shock at $x\approx 16\lambda_e$ in Fig.~\ref{figure05}(d). Some ambient ions were reflected by the shock forming the beam with the mean speed $v_x \approx 3.8 v_{fms}$. They trail a fast beam of rarefaction wave ions. The density overshoot at $x \approx 16 \lambda_e$ in Fig.~\ref{figure05}(e) reaches the density $3n_{i0}$. Its downstream plasma has the density $\approx 2n_{i0}$ because a subcritical shock heats ions only along the shock normal direction~\cite{Livadiotis2015}. Figure~\ref{figure05}(f) shows a magnetic overshoot. The magnetic field of the upstream plasma is compressed as it crosses the shock and its amplitude is larger than $B_0$ in its downstream region $11\lambda_e \le x \le 15\lambda_e$. The FMS shock is fully developed and planar.

\begin{figure*}
\includegraphics[width=\columnwidth]{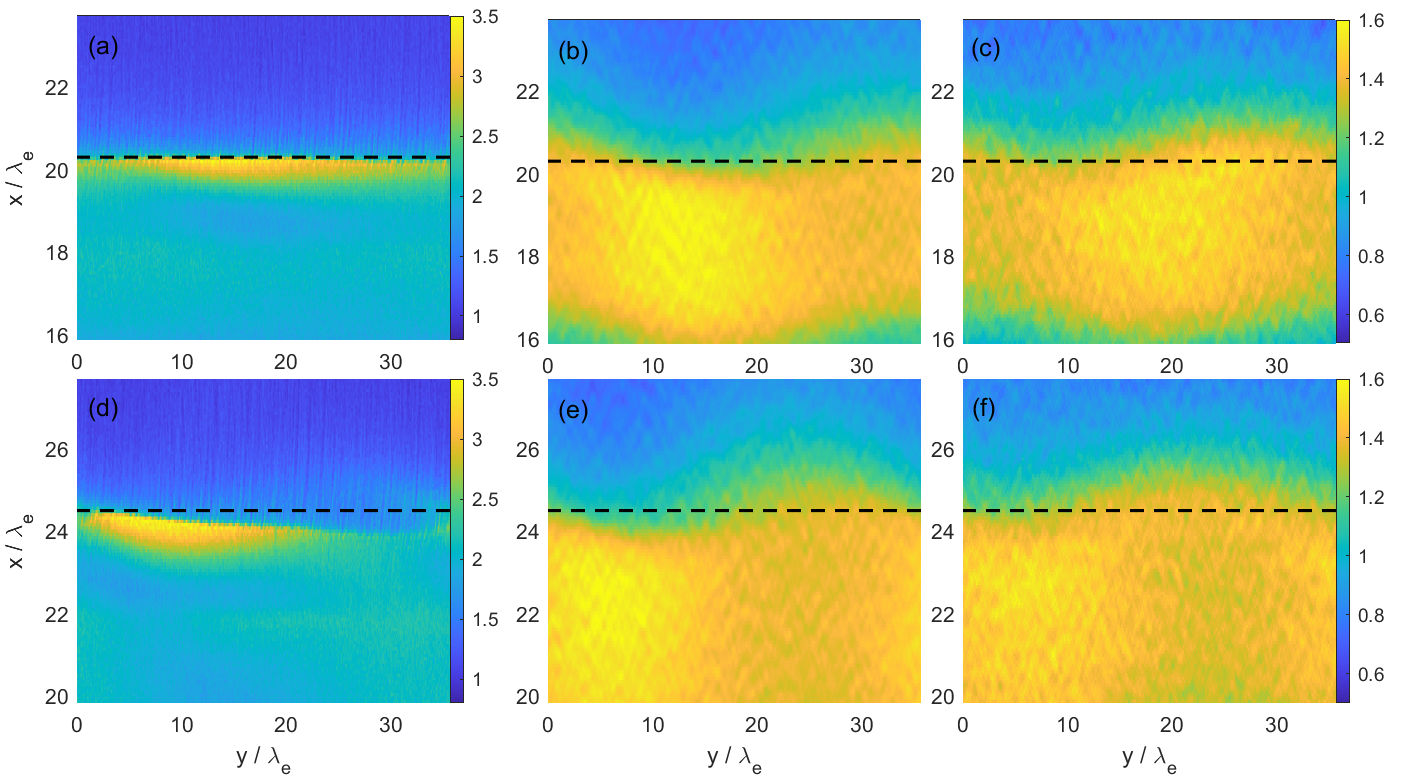}
\caption{Ion density and magnetic field distribution at $\omega_{lh}t=8$ (left column) and 10 (right column): Panels~(a,~d) show the ion density normalized to $n_{i0}$. Panels~(b,~e) show the distribution of $B_y/B_0$. Panels~(c,~f) show $B_z/B_0$. The vertical dashed lines mark $x=20.3\lambda_e$ in the left column and $x=24.5\lambda_e$ in the right column.}
\label{figure06}
\end{figure*}

Figure~\ref{figure06} shows the distributions of the ion density and magnetic field at the times $\omega_{lh}t = 8$ and 10. 
The shock front is no longer spatially uniform at $\omega_{lh}t=8$. While the density distribution in Fig.~\ref{figure06}(a) follows approximately the dashed line $x=20.3\lambda_e$, the density oscillates along y. Its maximum is located at $y\approx 18 \lambda_e$. Both magnetic field components have well defined fronts, which oscillate around the dashed black lines, and are shifted in phase. The $B_y$ component in Fig.~\ref{figure06}(b) and the $B_z$ component in Fig.~\ref{figure06}(c) extend farthest upstream at $y\approx 36\lambda_e$ and $y\approx 23\lambda_e$, respectively. A smaller phase shift is observed downstream of the shock at $x =  18 \lambda_e$, where the amplitude maxima of the magnetic field components peak at $y\approx 14\lambda_e$ and $y\approx 18\lambda_e$. The variation of the phase shift with x can be attributed to magnetic field compression by the shock, which is located where the density peaks in Fig.~\ref{figure06}(a). The rear boundary of the interval with the amplified magnetic at $x\approx 16\lambda_e$, which corresponds to the discontinuity separating the shock's downstream region from the rarefaction wave, oscillates in phase with the magnetic boundary near the shock front. 

At $\omega_{lh}t=10$, the front of the density overshoot is no longer planar. The density accumulation is concentrated in the interval $0 \le y/\lambda_e \le 20$ with its peak being located at $y\approx 10\lambda_e$. Its front is is tilted by about $1.4^\circ$ relative to y. The center of the density accumulation has propagated $\approx 8\lambda_e$ to decreasing y during the time interval $2/\omega_{lh}$ giving it the speed $\approx 1.6\times 10^6$ m/s or $5.7c_s$, where $c_s$ is the ion acoustic speed in the ambient plasma. A more accurate estimate is given below. The perturbation grew to a large amplitude between $\omega_{lh}t=5.8$ and 10, which suggests an exponential growth that is close to $\omega_{lh}$. The magnetic field oscillations in the interval $x < 24 \lambda_e$ in Figs.~\ref{figure06}(e,~f) are in phase and are correlated with the density oscillation near the dashed line. Their phase shift is about $10\lambda_e$ at the dashed line, which is close to the phase shift $13\lambda_e$ at $\omega_{lh}t=8$.

\begin{figure*}
\includegraphics[width=\columnwidth]{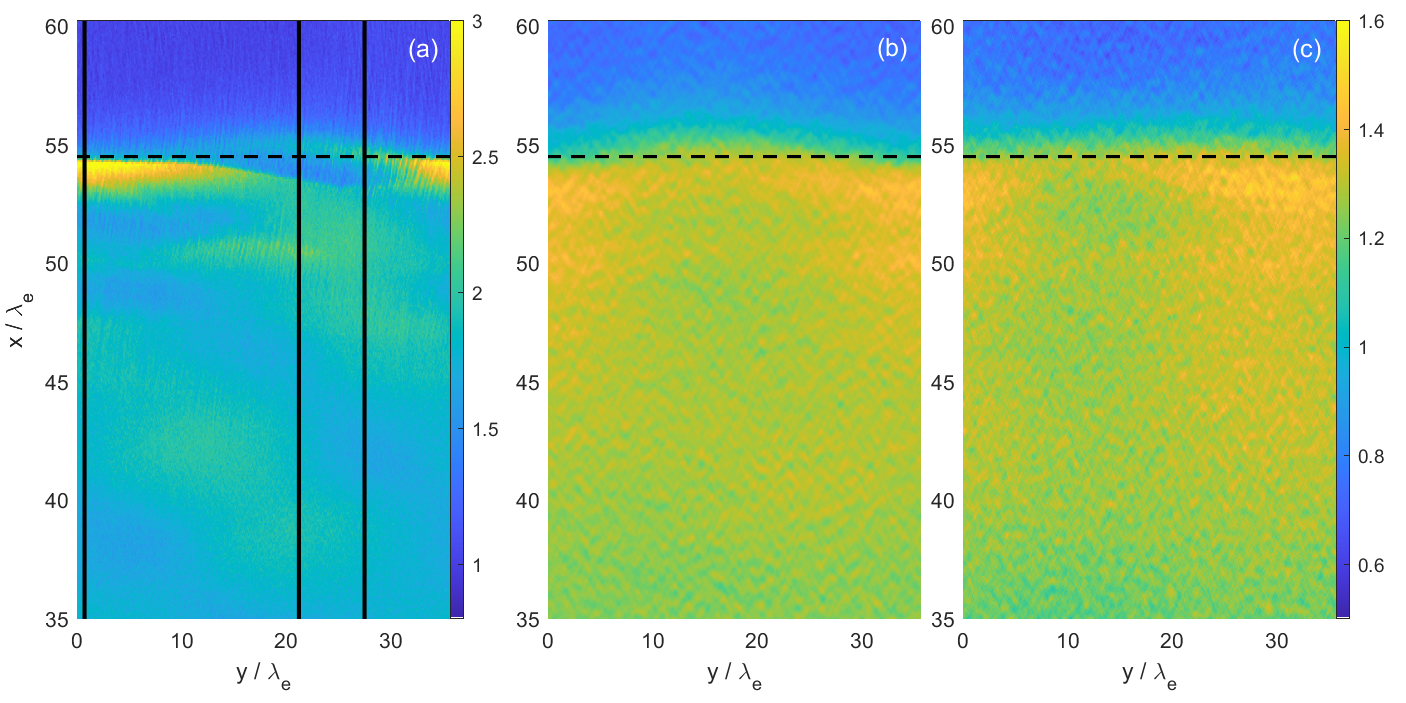}
\caption{The distributions of the ion density and magnetic amplitudes at $\omega_{lh}t=25$: Panel~(a) shows the ion density in units of $n_{i0}$. The 3 horizontal solid lines show the positions $y/\lambda_e=$ 0.7, 21, and 27.5. Panels~(b,~c) show $B_y/B_0$ and $B_z/B_0$. The vertical dashed lines mark $x=54.5\lambda_e$.}
\label{figure07}
\end{figure*}

Figure~\ref{figure07} shows the distributions of the ion density and the amplitudes of $B_y$ and $B_z$ at the time $\omega_{lh}t = 25$.
The shock front, indicated by the dashed line, has propagated $30\lambda_e$ during the time $15/\omega_{lh}$ giving the speed $1.65v_{fms}$. The ion density near the front is still varying in Fig.~\ref{figure07}(a). A density modulation is also visible downstream of the shock. For example, the density at the lineout $x=42\lambda_e$ oscillates between $2n_{i0}$ at $y\approx 10\lambda_e$ and $1.65n_{i0}$ at $y\approx 30\lambda_e$. The density oscillation drifts to increasing y with increasing distance from the shock front. This drift is caused by a nonuniform compression of the plasma by a shock modulation that drifts to decreasing y.

The magnetic field components still show a surface wave near $x=54.5\lambda_e$ and both components are shifted in phase. Magnetic field compression by the shock boundary, which is modulated along y, results in a variable compression of the downstream magnetic field. We observe amplitude modulations of both components along y that are not shifted in phase in the interval $40 \le x / \lambda_e \le 53$.   

\begin{figure*}
\includegraphics[width=\columnwidth]{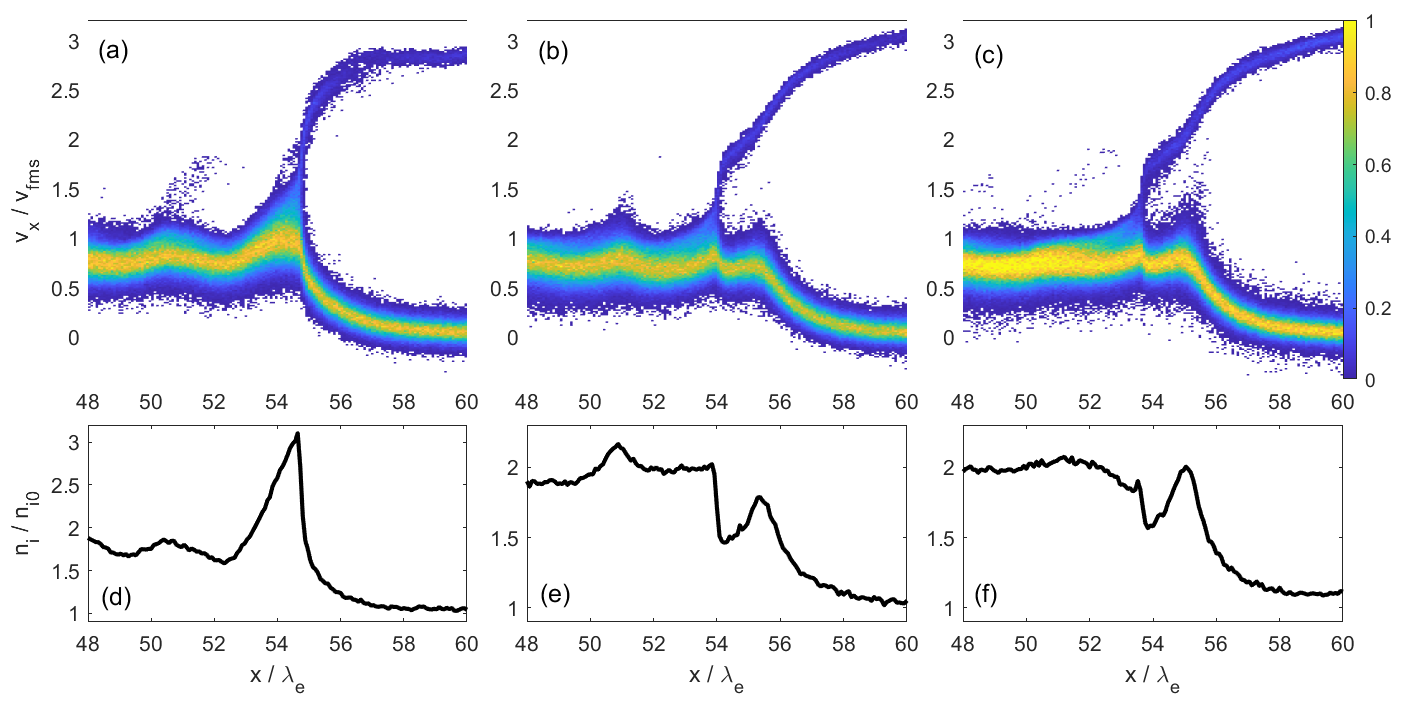}
\caption{Panels~(a-c) show the phase space density distributions of ions at the positions $y/\lambda_e=$ 0.7, 21, and 27.5. The time is $\omega_{lh}t=25$. All densities are normalized to the maximum value of the phase space density of the ambient plasma at the time $t=0$ and displayed on the same linear color scale. Panels~(d-f) show the ion densities, in units of $n_{i0}$, corresponding to the phase space density above.}
\label{figure08}
\end{figure*}

Figure~\ref{figure08} examines the phase space density distributions of the ions in the $x, v_x$ slice plane along the three vertical lines shown in Fig.~\ref{figure07}(a). The phase space density is averaged over slices 45 grid cells wide along y that are centered on the displayed lines. The figure also shows the ion density averaged over these y intervals.
Figures~\ref{figure08}(a,~d) show the ion distribution close to $y=0.7\lambda_e$. The phase space density shows a structure resembling that of the shock in Fig.~\ref{figure05}(d) located at $x\approx 54.5$ coinciding with that of the peak of the density overshoot. The overshoot is separated from a second peak downstream at $x\approx 50.5\lambda_e$. Their separation is 30\% larger than that in Fig.~\ref{figure04}, which is probably caused by the interplay of the lower-hybrid mode that mediates the shock and the oblique Whistler wave, which perturbs it. 

The ion distribution in the slice centered on $y=21\lambda_e$ shows that the ambient ions are accelerated already at $x\approx 55.5\lambda_e$ reaching the speed $0.7v_{fms}$. The shock-reflected ion beam merges with the accelerated ambient ions at $x\approx 54\lambda_e$ where the ion density rises in Fig.~\ref{figure08}(e) but remains well below that of the overshoot in Fig.~\ref{figure08}(d). This is what is left of a collapsed FMS shock that was convected downstream. At this time, the front of the accelerated ions corresponds to a magnetic piston. A magnetic piston is mediated by the Hall electric field of the moving magnetic field. An FMS shock combines this field with the thermoelectic field of the density overshoot. The magnetic piston in Fig.~\ref{figure08}(c) is causing the steepening of an FMS wave near $x=55\lambda_e$, which lets the density overshoot grow in Fig.~\ref{figure08}(f). Eventually this structure will become a new FMS shock. 

The propagation of the oblique Whistler wave along the magnetic overshoot leads to a periodic collapse and reformation of the subcritical FMS shock on time scales $\omega_{lh}^{-1}$. Figure~\ref{figure08} shows that the ion density is higher behind the magnetic piston than behind the FMS shock. Figure~\ref{figure07}(a) supports this observation showing that the ion density downstream is largest between the lines $x/\lambda_e = 21$ and 27.5.

The consistent phase shift between the amplitudes of $B_y$ and $B_z$ in the magnetic overshoot suggests that we can use these components to determine additional wave properties. The background magnetic field is $\mathbf{B}_0=(0,-B_0,B_0)/\sqrt{2}$. As long as the magnetic field is not rotated, the sum of the $B_y$ and $B_z$ components gives zero. A non-zero value of $\Delta_B(x,y) = B_y(x,y)+B_z(x,y)$ thus indicates a rotation of the magnetic field. We sample $\Delta_B(x,y)$ at a position $x^*$, which is constant in the shock frame moving with the speed $1.68v_{fms}$. This position equals that of the dashed horizontal lines in Fig.~\ref{figure06} and Fig.~\ref{figure07}. We also write out the ion density at this lineout.

\begin{figure*}
\includegraphics[width=\columnwidth]{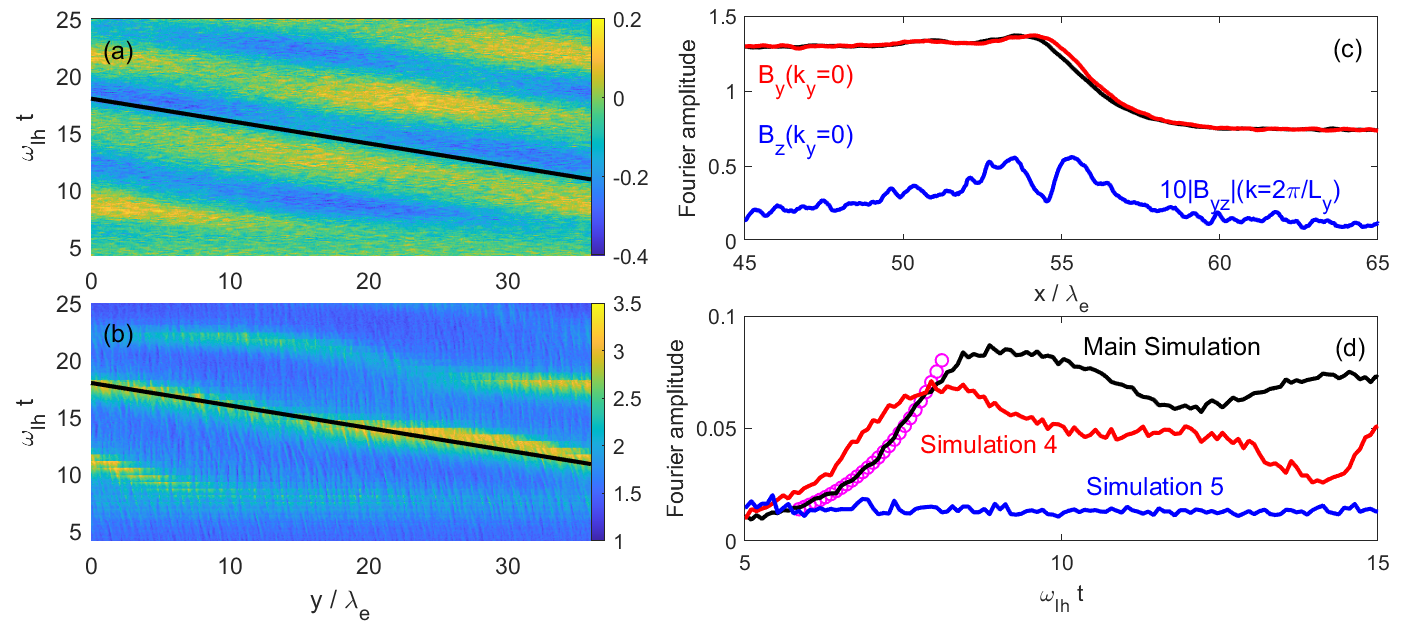}
\caption{Panel~(a) shows $\Delta_B(x^*,y)$ at the position $x^*$, which is moving with the FMS speed $1.68v_{fms}$ and is positioned in the magnetic overshoot ahead of the shock.  Panel~(b) shows the ion density along the same lineout. The black lines correspond to the speed $0.4v_D$ with the electron drift speed $v_D=-5.3\times 10^6$m/s. Panel~(c) shows the moduli of the Fourier amplitudes of $B_y$ and $B_z$ at the wavenumber $k_y=0$ and time $\omega_{lh}t=25$ in the main simulation. It also shows the modulus $10|B_{yz}|$ of the Fourier amplitude of $B_y + i B_z$ at the wavenumber $k_y=2\pi/L_y$. Panel~(d) shows the evolution of the largest Fourier amplitude modulus of $B_y$ near the shock for the smallest resolved wavenumber $k_y$. This wavenumber is $k_y=2\pi/L_y$ in the main simulation, $k_y = 2.66\pi/L_y$ in simulation 4 and $k_y = 4\pi / L_y$ in simulation 5. The magenta circles follow an exponential growing at the rate $\omega_{lh}/1.3$.}
\label{figure09}
\end{figure*}

Figure~\ref{figure09}(a) shows a magnetic perturbation that grows after $\omega_{lh}t=6$, propagates at the phase speed $-0.4|v_D|$, and oscillates around its equilibrium after saturating.
Its phase speed is $0.6|v_D|$ or $6.3v_{fms}$ in the rest frame of the electrons matching that of the oblique Whistler wave in Fig.~\ref{figure03}(d) at $k_y = 2\pi /L_y$. The oblique Whistler wave, with its phase velocity $-2.1 \times 10^6$~m/s in the rest frame of the simulation box, is faster than the ion density wave with the phase speed $c_s$. Both waves are not in resonance confirming the finding of Ref.~\cite{Ji2005} that the instability is reactive or fluid-like. Figure~\ref{figure09}(b) shows a density perturbation, which follows the magnetic perturbation. This perturbation is neither stable nor sinusoidal but has steepened density profile similar to an electrostatic shock, which is probably a consequence of its propagation speed $\gg c_s$.

According to Ref.~\cite{Ji2005}, waves grow as long as the phase speed of the backward-propagating oblique Whistler wave, measured in the rest frame of the simulation box, is faster than the ion acoustic wave that propagates in the direction of the drifting electrons. This is the case for $k_y=2\pi/L_y$ in Fig~\ref{figure03}(d), where the drift speed modulus exceeds the phase speed of the oblique Whistler wave. The first harmonic $k_y=4\pi/L_y$ is already too fast to induce wave growth. Since its phase speed exceeds $|v_D|$, it propagates backward even in the box frame and does therefore not outrun the ion acoustic wave. 

Simulations~4 and~5 resolve intervals along y with the respective lengths $3L_y/4$ and $L_y/2$. Equation~\ref{whistler} gives the phase speed $8.4v_{fms}$ or $4.3\times 10^6$~m/s for the oblique Whistler wave with $k_y = 2.66\pi/L_y$. The backward propagating oblique Whistler wave has the phase speed $-10^6$~m/s or $-3.5c_s$ in the box frame outrunning the ion acoustic wave that propagates in the negative y direction, which should lead to the growth of an oblique Whistler wave in simulation~4. This condition is not met in simulation~5.  

We test wave growth in simulations~4 and~5 by examining their amplitude in time. Figure~\ref{figure09}(c) illustrates the method. We Fourier-transform the distributions of $B_y$ and $B_z$ in the left half of Figure~\ref{figure01} over y. We plot the amplitude moduli of $B_y$ and $B_z$, which are both normalized to $B_0$, for $k_y=0$. Both have amplitudes $1/\sqrt{2}$ ahead of the shock and twice that behind the shock. We also show the modulus of the Fourier transform of $B_y+iB_z$ for $k_y=2\pi/L_y$. Its amplitude peaks in the shock transition layer. 

We Fourier transform $B_y$ over y in the main simulation and in simulations~4 and~5 and extract the largest amplitude of the smallest resolved wavenumber near the shock. Their time series are shown in Fig.~\ref{figure09}(d). In the main simulation, the amplitude starts to grow at the exponential rate $\omega_{lh}/1.3$ for $6 \le \omega_{lh}t \le 7.5$. After the oblique Whistler wave has saturated, it performs oscillations at the frequency $\approx \omega_{lh}$. The oblique Whistler wave in simulation~4 starts to grow first. Its exponential growth rate is comparable to that in the main simulation but it saturates at a lower amplitude. The faster growth might be caused by a smaller mismatch of the phase velocities of the oblique Whistler wave and the ion acoustic wave, while the lower saturation amplitude might be connected to stronger electron cyclotron damping. Simulation~5 shows that the oblique Whistler wave with the wavenumber $k_y = 4\pi /L_y$ is indeed stable.

\section{Perturbed shock}

\begin{figure*}
\includegraphics[width=\columnwidth]{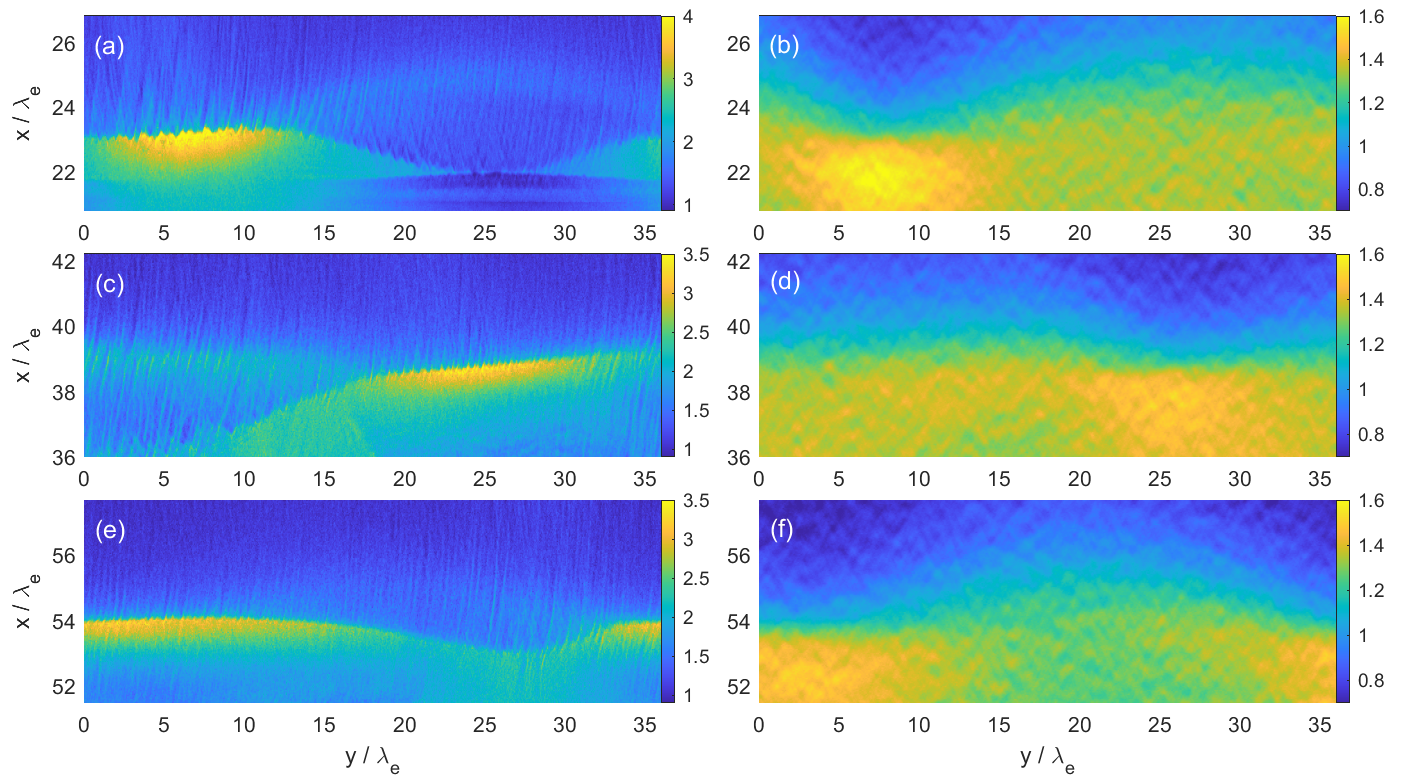}
\caption{The left column shows the ion density near the shock in units of $n_{i0}$. The right column shows $B_y/B_0$. Panels~(a) and (b) correspond to the time $\omega_{lh}t=10$. Panels~(c,~d) correspond to $\omega_{lh}t=17.5$ and (e,~f) to $\omega_{lh}t=25$.}
\label{figure10}
\end{figure*}

We discuss the evolution of the shock that propagates to the right in Fig.~\ref{figure01}. Figure~\ref{figure10} shows the distributions of the ion density and the magnetic $B_y$ component near the shock at three times.
Panels~(a,~b) show both distributions after the shock left the perturbation layer, which has its boundary at $x=20.8\lambda_e$. The number density of mobile ions was largest at $y=9\lambda_e$ and lowest at $y=27\lambda_e$. A density overshoot is visible in Fig.~\ref{figure10}(a) only for $2 \le y / \lambda_e \le 13$. Its thickness $\sim \lambda_e$ is comparable to that in Fig.~\ref{figure06}(d). This density overshoot and the compressed magnetic field behind it in Fig.~\ref{figure10}(b) demonstrate that this is a FMS shock. 

Hardly any plasma compression takes place near $y=27\lambda_e$. The magnetic field bulges out in the upstream direction and the contour lines $B_y\approx 1.2B_0$ correlate with an ion density structure. A FMS wave is growing in response to the Hall electric field induced by the drifting compressed magnetic field. This FMS wave changed into an FMS shock in Figs.~\ref{figure10}(c,~d) again characterized by a density overshoot and by the compression of the magnetic field that crosses it. As this FMS shock formed, the one at the earlier time collapsed; the FMS shock and the magnetic piston have swapped places. A density enhancement with $x\approx 39\lambda_e$ and $y \le 12\lambda_e$ in Fig.~\ref{figure10}(c) evidences the growth of a new FMS wave. It turned into a shock at the simulations end. An FMS shock is located in Figs.~\ref{figure10}(e,~f) at $x\approx 54\lambda_e$. Its left boundary is near $y \le 34 \lambda_e$, it crosses the periodic boundary and ends near $y=9\lambda_e$. A magnetic piston is visible near $y\approx 27\lambda_e$.  

\begin{figure*}
\includegraphics[width=\columnwidth]{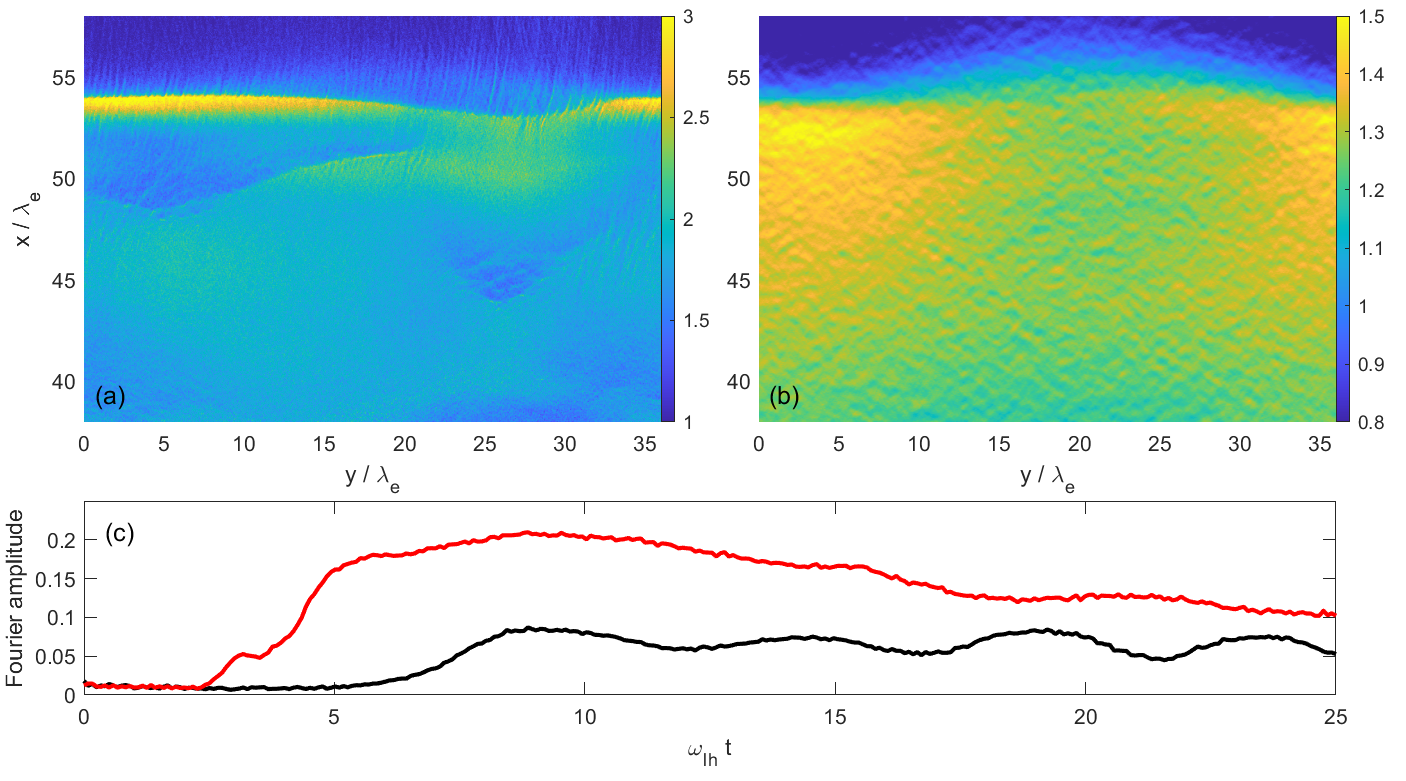}
\caption{Panel~(a) shows the ion density near the shock in units of $n_{i0}$ and~(b) shows $B_y/B_0$ at $\omega_{lh}t=25$. Panel~(c) compares the Fourier amplitudes of $B_y$ at $k_y=2\pi/L_y$ for the perturbed and unperturbed shocks normalized as in Fig~\ref{figure09}(d).}
\label{figure11}
\end{figure*}

Figures~\ref{figure11}(a,~b) show the ion density and magnetic $B_y$ component over a large interval in $x$.
These figures reveal the extent of the density and magnetic field oscillations caused by the perturbed shock. The number density behind the magnetic piston near $x=54\lambda_e$ and $y=27\lambda_e$ is larger than behind the FMS shock near $y=9\lambda_e$. The density distribution is reversed near $x=46\lambda_e$; Figure~\ref{figure10}(c) shows that the FMS shock and the magnetic piston had their places swapped at this time. The periodic reformation of the shock thus causes density modulations behind the shock that extend at least 10$\lambda_e$ downstream of it. The density depletion at $x\approx 46\lambda_e$ and $y=27\lambda_e$ is not correlated with a magnetic enhancement. However, the $B_y$ component shows higher amplitudes behind the density overshoot that extend about $7\lambda_e$ downstream. Since the density overshoot exchanges its position with the magnetic piston, this magnetowave is a standing wave that oscillates at the frequency of the shock reformation. Only one reformation is observed between $\omega_{lh}t=10$ and $\omega_{lh}t=25$ giving a frequency $\approx \omega_{lh}/2$. The absence of a propagating oblique Whistler wave near the shock suggests that these waves can only grow if the shock front is planar to start with.  

The frequency with which the shock reforms is well below that of the oblique Whistler wave $1.28\omega_{lh}$ at the wavenumber $k_y =2\pi/L_y$ and we can therefore not observe any coupling between the shock oscillation and the oblique Whistler wave. In Ref.~\cite{Dieckmann2023}, the magnetic field with amplitude $B_0$ pointed along y. The in-plane magnetic field, which provides the restoring force for the shock boundary oscillation, was thus $\sqrt{2}$ times stronger than the one in the main simulation and gave rise to a shock reformation frequency that was higher by the same factor. This corroborates that magnetic tension is responsible for the shock boundary oscillations.

Figure ~\ref{figure11}(c) compares the maximum values of the Fourier amplitudes of the $B_y$ components in the main simulation for the domain with the unperturbed shock domain in the left of Fig.~\ref{figure01} with that in the right with the perturbed shock. For the latter simulation, the amplitude starts to increase at $\omega_{lh}t=2.5$, when the FMS shock enters the perturbation layer, and reaches its peak value just before $\omega_{lh}t=10$ when it is leaving it. The amplitude decreases continuously after this time demonstrating that the shock oscillation is damped. This is not the case for the wave propagating along the initially unperturbed shock where the amplitude oscillates around an equilibrium value.

\section{Discussion}

We investigated mechanisms that trigger surface oscillations of subcritical perpendicular fast magnetosonic shocks. Our study extends previous ones~\cite{Dieckmann2023,Dieckmann2024,Dieckmann2025} keeping all plasma conditions unchanged compared to these previous studies except the orientation of the background magnetic field relative to the normal of the 2D simulation plane. The term shock surface refers to its magnetic overshoot. If the direction of the magnetic field in the upstream plasma was resolved by the simulation plane, no instabilities grew and an external perturbation of the shock surface led to stable surface oscillations~\cite{Dieckmann2023}. They were caused by a cyclic reformation of the subcritical shock. Unlike the well-known cyclic reformation of supercritical shocks on time scales $\sim \omega_{ci}^{-1}$, the oscillations involve the much faster timescales $\omega_{lh}^{-1}$. They depend on the coupling of adjacent FMS shocks and magnetic pistons by magnetic tension. A magnetic piston that turns into an FMS shock lets the neighboring FMS shock collapse. If the background magnetic field was aligned with the simulation plane normal such oscillations could not develop due to the absent magnetic tension. However, the electron drift current led to an electron-cyclotron drift instability ahead of the shock's density overshoot and to a lower-hybrid drift instability behind it.  

Here we rotated the background magnetic field by $\theta = 45^\circ$ relative to the simulation plane normal but kept it perpendicular to the shock normal. We observed the growth of oblique Whistler waves in the shock surface ahead of the density overshoot. Such waves are frequently observed in simulations of supercritical magnetized shocks. It is well-known that their energy source is the diamagnetic electron current but, to the best of our knowledge, the exact instability mechanism remained elusive. The modified two-stream instability, which can generate oblique Whistler waves ahead of supercritical quasi-perpendicular FMS shocks, is ineffective if the oblique Whistler waves have a wave vector that is not almost perpendicular to the upstream magnetic field. 

Another instability mechanism is the oblique lower-hybrid drift instability proposed by~\cite{Ji2005}. The authors considered a Harris-type equilibrium, which is relevant for magnetic reconnection experiments. They employed a magnetized fluid model for the electrons while the drifting ions were considered to be unmagnetized. Waves grow in this model by the reactive coupling of the backward-propagating oblique Whistler wave with the ion acoustic wave. The wavenumber spectrum of unstable waves was derived using this simplified model as well as the growth rates of the oblique Whistler waves. The analytic model in Ref.~\cite{Ji2005} was developed for a stationary plasma, in which the thermal pressure change is balanced by a magnetic pressure change. 

In the shock's magnetic overshoot ahead of the density overshoot, the magnetic pressure change is sustained by the ram pressure of the upstream plasma. Although the ions are slowed down as they cross the shock transition layer, they still move at a speed along the shock normal that exceeds $v_{fms}$. However, since the phase speed of the shock boundary wave exceeds by far the shock speed in the upstream frame, the Harris sheet and shock scenarios remain similar.

By employing periodic boundary conditions along the direction of the wavevector, individual wave numbers can be isolated in PIC simulations.  Our main simulation and simulations~4 and~5 used box lengths that resolved the oblique Whistler waves near the maximum unstable wavenumber predicted by~\cite{Ji2005}. The simulations
demonstrated that the oblique Whistler waves no longer grow if their wavelength approaches the wave number at which the waves stabilize in the analytic model.

The phase velocity of the surface wave was that of the oblique Whistler wave and exceeded by far the ion acoustic speed. Its supersonic speed let the ion density wave steepen into a shock and the amplitude of its density oscillations was comparable to the amplitude of the shock's density overshoot. Since the presence of a density overshoot is the key difference between an FMS shock and a magnetic piston, the oblique Whistler wave triggered a cyclic reformation of the FMS shock. It became phase-correlated with the state of the shock. Unlike the cases discussed in Refs.~\cite{Dieckmann2023,Dieckmann2024,Dieckmann2025}, the cyclic reformation did not have to be triggered by an external perturbation.

Our simulations showed that the oblique Whistler wave grew at an exponential rate that exceeded the one predicted from linear theory by the factor 8. This discrepancy might be caused by the thermally anisotropic electron distribution in the shock transition layer, by a density profile that was not sinusoidal or by the phase correlation of the wave with the FMS shock structure. 

The most likely cause is the thermally anisotropic electron distribution. In the Harris-type configuration in~\cite{Ji2005}, the electrons had a thermally isotropic velocity distribution giving a temperature that does not depend on the direction. A shock compresses the magnetic field, which accelerates electrons orthogonally but not parallel to the magnetic field. Such a distribution triggers the growth of Whistler waves. The interplay between the reactive instability and the thermally anisotropic electron distribution may be responsible for the accelerated wave growth.

According to~\cite{Ji2005}, the spectrum of waves, which grow in response to the oblique lower-hybrid drift instability, and their growth rates depend on the electron inertia but the waves grow even if the electron inertia is zero. This aspect may explain why oblique Whistler waves were observed in PIC simulations and in hybrid simulations, which consider an inertialess electron fluid~\cite{Lembege2009}. However, given that these shocks were supercritical, the instability mechanism may differ from the one discussed here.

If the shock boundary is initially planar, the oblique lower-hybrid drift instability drives waves with the hybrid wavelength $\lambda \sim (\lambda_i \lambda_e)^{1/2}$ and frequencies close to the lower-hybrid frequency. However, if the waves propagate along the shock surface at an angle $\theta < 45^\circ$, the longest unstable wavelengths can become significantly larger than $\lambda_i$ bringing them closer to MHD scales. In this case, the shock curvature limits $\lambda$ since the waves can only grow if the magnetic overshoot is sufficiently planar. 

Supercritical perpendicular FMS shocks have a thicker magnetic overshoot. Both, subcritical and supercritical FMS shocks, do however have a density overshoot with a similar thickness. Since the oblique Whistler waves have growth rates $\approx \omega_{lh}\gg \omega_{ci}$, they may grow and saturate before the supercritical shock collapses. Indeed, some PIC simulations found this to be the case and observed that the oblique Whistler waves can suppress the shock reformation. Future work has to assess the degree to which the oblique lower-hybrid instability can let oblique Whistler waves grow at perpendicular supercritical shocks and if the waves could be the reason for shock rippling by the MMS mission. 

Future work must also determine if the thermal anisotropy of the electrons is responsible for the larger exponential growth rate of the oblique Whistler waves. If this is the case, the reactive instability proposed by~\cite{Ji2005} is not only important in the context of magnetic reconnection but also for the stability of subcritical and possibly also supercritical shocks.

\section*{Acknowledgements} The simulations were performed on resources provided by the National Academic Infrastructure for Supercomputing in Sweden (NAISS) at the National Supercomputer Centre partially funded by the Swedish Research Council through grant agreement no. 2022-06725 and on the centers of the Grand Equipement National de Calcul Intensif (GENCI) under grant number A0150406960.

\section*{Data availability statement} All data that support the findings of this study are included within the article (and any supplementary files).

\section*{Conflict of interest} The authors declare that they have no conflict of interest.

\end{document}